\documentclass[aps,prb,onecolumn,superscriptaddress,floatfix,showpacs,amsmath,amssymb,nofootinbib,longbibliography,nobibnotes,noeprint,11pt]{revtex4-2}

\usepackage{graphicx}% Include figure files
\usepackage{dcolumn}% Align table columns on decimal point
\usepackage{bm}% bold math
\usepackage{color}
\usepackage{tabularx}
\usepackage{csquotes}
\usepackage[normalem]{ulem}
\usepackage{xcolor}

\newcommand{\rbwo}{$R_3$BWO$_9$}
\newcommand{\pbwo}{Pr$_3$BWO$_9$}

\newcommand{\nbwo}{Nd$_3$BWO$_9$}

\newcommand{\lbwo}{La$_3$BWO$_9$}

\newcommand{\prion}{Pr$^{3+}$}

\newcommand{\mbpr}{\(\mu _{\rm B}/{\rm Pr^{3+}}\)}
\newcommand{\mb}{\(\mu _{\rm B}\)}

\newcommand{\jmk}{J/(mol$_{\rm Pr}$\,K)}

\newcommand{\cpp}{$c_{\rm p}$}
\newcommand{\cpel}{$c^{\rm el}_{\rm p}$}

\begin{document}

\title{Magnetic and thermodynamic studies on the distorted kagome magnet Pr$_3$BWO$_9$}

%%%%  AUTHORS %%%%%%%%%%%%%%%
\author{A. Elghandour}
\affiliation{Kirchhoff Institute for Physics, Heidelberg University, INF 227, D-69120 Heidelberg, Germany}
\affiliation{European XFEL, D-22869 Schenefeld, Germany}

\author{J.~Arneth}
\affiliation{Kirchhoff Institute for Physics, Heidelberg University, INF 227, D-69120 Heidelberg, Germany}

\author{A. Yadav}
\affiliation{Department of Physics, Indian Institute of Technology Madras, Chennai 600036, India}

\author{S.~Luther}
\affiliation{Dresden High Magnetic Field Laboratory (HLD-EMFL), Helmholtz-Zentrum Dresden-Rossendorf, 01328 Dresden, Germany}

\author{P. Khuntia}\email{pkhuntia@iitm.ac.in}
\affiliation{Department of Physics, Indian Institute of Technology Madras, Chennai 600036, India}
\affiliation{Quantum Centre of Excellence for Diamond and Emergent Materials, Indian Institute of Technology Madras, Chennai 600036, India}

\author{R.~Klingeler}
\email{Corresponding author: klingeler@kip.uni-heidelberg.de}
\affiliation{Kirchhoff Institute for Physics, Heidelberg University, INF 227, D-69120 Heidelberg, Germany}

\date{\today}

\begin{abstract}
We report specific heat, ac/dc magnetic susceptibility as well as static and pulsed field magnetization studies on the distorted kagome magnet Pr$_3$BWO$_9$ down to 0.4~K and up to high magnetic fields. The low-temperature thermodynamic properties are found to be governed by an electronic quasi-doublet ground state; the energy splitting of which amounts to $\Delta_1\simeq 18$~K and exhibits a quadratic field dependence with $g_\mathrm{eff} = 2.6$. Fitting of the specific heat data implies that the next excited state is strongly gapped at $\Delta_2=430$~K and three-fold degenerate in zero field. Our dc and ac susceptibility studies down to 0.4~K do not detect signatures of distinct spin glass behavior. Pulsed field magnetization measurements up to 60~T confirm the Ising-like paramagnetic nature of the magnetic ground state which is characterized by $m_J=\pm 4$ and the anisotropy energy $E_a\simeq 950$~K.
\end{abstract}
\maketitle

%%%%%%%%%%%%%%%%%%%%%%%%%%
\section{Introduction}
%%%%%%%%%%%%%%%%%%%%%%%%%%

Frustrated quantum magnets, in which a delicate interplay among emergent degrees of freedom is active, serve as excellent candidate materials for hosting a wide array of exotic many-body phenomena that transcend conventional paradigms~\cite{balents2010spin,broholm2020,Zhou2017,Savary2012,Savary2016,Khatua2023}. In lanthanide-based materials with strong geometric frustration, a rich variety of exotic magnetic ground states can be studied by varying between the Heisenberg and Ising limits, and between the quasi-classical and quantum limits, through appropriate choice of the rare earth ions. Due to strong intrinsic spin-orbit coupling in the 4$f$ shell and often dominant dipolar magnetic interactions, rare earth-hosting systems avoid perturbations found in 3$d$ transition metal-based candidate materials and, hence, hold promise to realize novel spin liquid states. The localized nature of the 4$f$ electrons typically leads to weak magnetic interaction energies, such that spin dynamics associated with geometrical frustration often emerge at low temperatures far below the first excited crystal field level, while rather moderate, i.e., accessible, magnetic fields are sufficient to cover the full energy range. Prominent examples of such material families are pyrochlores $R_2$Ti$_2$O$_7$~\cite{ramirez1999zero,bramwell2001spin,Savary2012,Paddison2021,Elghandour2024}, filled hexagonal lattice indates $R$InO$_3$~\cite{clark2019,yuan2023},
langasites $R_3$Ga$_5$SiO$_{14}$~\cite{Zhou2007,simonet2008hidden,Weymann2020}, triangular NdTa$_7$O$_{19}$~\cite{Arh2021}, or the tripod kagome magnets $R_3$Mg$_2$Sb$_3$O$_{14}$~\cite{Dun2016,Paddison2016,khatua2022magnetic,Wellm2020}.

\begin{figure} [htb] 
    \includegraphics[width = \columnwidth]{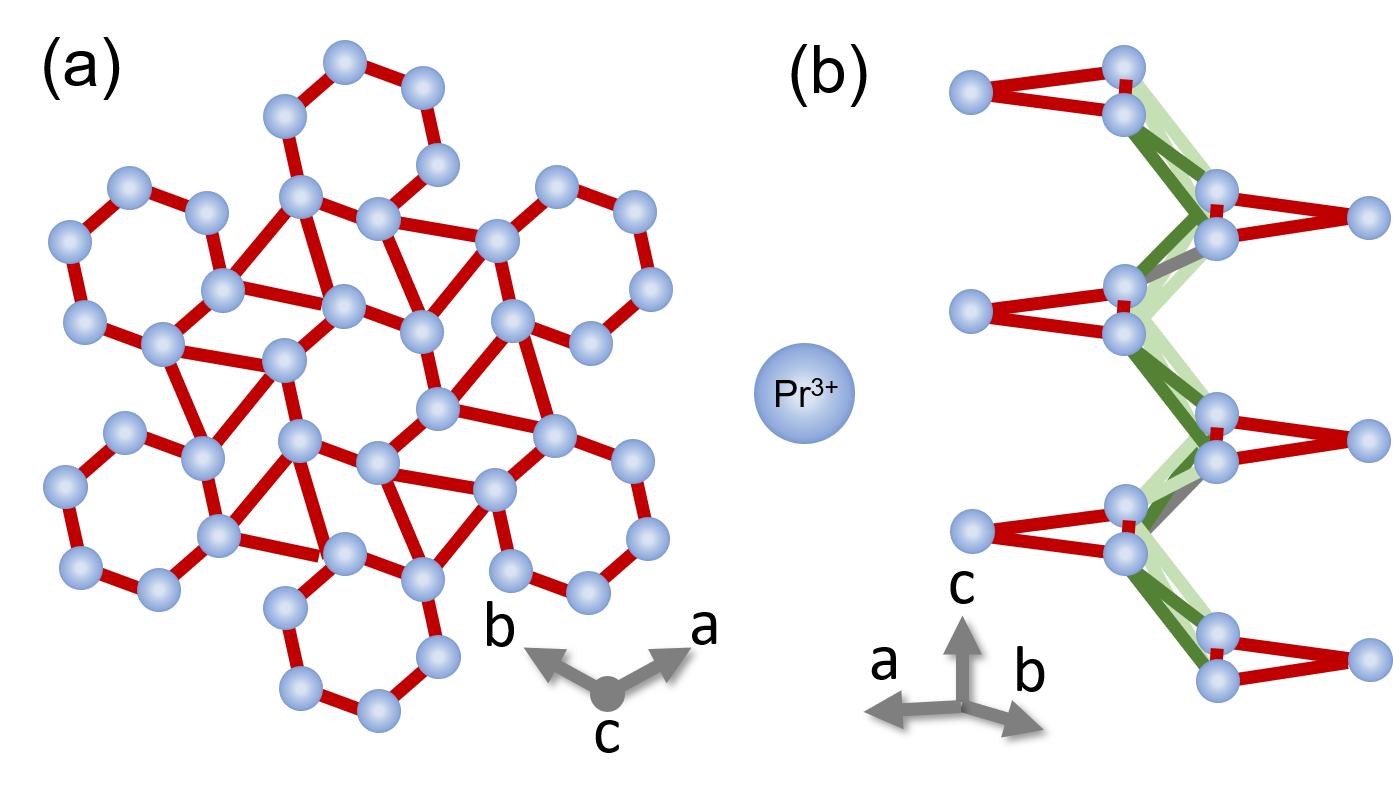}
    \caption{Schematics of the \prion\ sites (blue spheres) in \pbwo. (a) In-plane ($ab$ plane) arrangement and (b) stacking along the $c$ axis. Lines indicate short in-plane (red) and interplanar (green) Pr-Pr distances.}
    \label{unitcell}
\end{figure}

Boratotungstates \rbwo\ have recently entered the stage as a new family of frustrated quantum magnets~\cite{ashtar2020new,yang2020large,zeng2021local,zeng2022incommensurate,Flavi,Nagl2024,Yadav2025}. Among them, \pbwo\ in particular exhibits no long-range magnetic order down to at least 90~mK~\cite{zeng2021local,Nagl2024}. The material crystallizes in a hexagonal coordinated structure with space group $P6_{\mathrm 3}$ %(no.~173), 
where the paramagnetic \prion\ ions form a distorted kagome lattice within the $ab$ planes which are stacked in an AB-type fashion along the $c$ axis~\cite{ashtar2020new,zeng2021local,Flavi} (see Fig.~\ref{unitcell}). 
The distorted kagome lattice features triangles of two different Pr-Pr bond lengths with 4.312~\AA\ and 4.886~\AA , respectively, arranged in an alternating pattern. 
Two recent studies have investigated the magnetic ground state as well as low-energy magnetic excitations. A recent NMR study~\cite{zeng2021local} shows a persistent cooperative paramagnetic state at lowest temperatures and suggests a spin gap opening at the antiferromagnetic wave vector, the size of which linearly increasing in external magnetic fields. These results are contrasted by inelastic neutron scattering, heat capacity, ultrasound and muon spin relaxation measurements, which suggest a quantum disordered ground state~\cite{Nagl2024}. The observed data are  explained in terms of dispersive crystal field excitations partially broadened by weak structural disorder, and, hence, suggest that \pbwo\ is not a perfect kagome magnet but realizes a complex exchange topology of twisted triangular spin-tubes, moderately coupled by the frustrated planar interactions. 
 
Here, we report detailed specific heat as well as dc and ac magnetic studies on \pbwo\ up to high magnetic fields. No evidence of long-range order nor of glassy behavior is observed in our dc and ac magnetization studies down to 400~mK. The magnetic ground state is further investigated by magnetization studies up to 60~T confirming its Ising-like paramagnetic nature. In addition to the anisotropy energy $E_a\simeq 82$~meV, we determine the energy levels of the five lowest lying states of \prion\ and investigate the field dependence of the smallest excitation gap. %Our analysis by means of a multi-level model also yields an estimate of the Weiss temperature. Quantitatively, we find $\Theta \simeq -20$~K which suggests a frustration ratio $f = |\Theta|/T_{\mathrm{N}} \geq 220$ and confirms \pbwo\ as a highly frustrated Ising-like magnet.

%%%%%%%%%%%%%%%%%%%%%
\section{Experimental Methods}
%%%%%%%%%%%%%%%%%%%%%

%Polycrystalline samples of \pbwo\ were synthesized by a solid-phase reaction from Pr$_2$O$_3$ (Alfa Aesar, 99.999~\%), as starting materials following the procedure in \cite{ashtar2019synthesis,Krutko2006} and similar as described previously by us for \nbwo~\cite{Yadav2025}.

\begin{figure}[htb]
    \centering
    \includegraphics[width=\columnwidth]{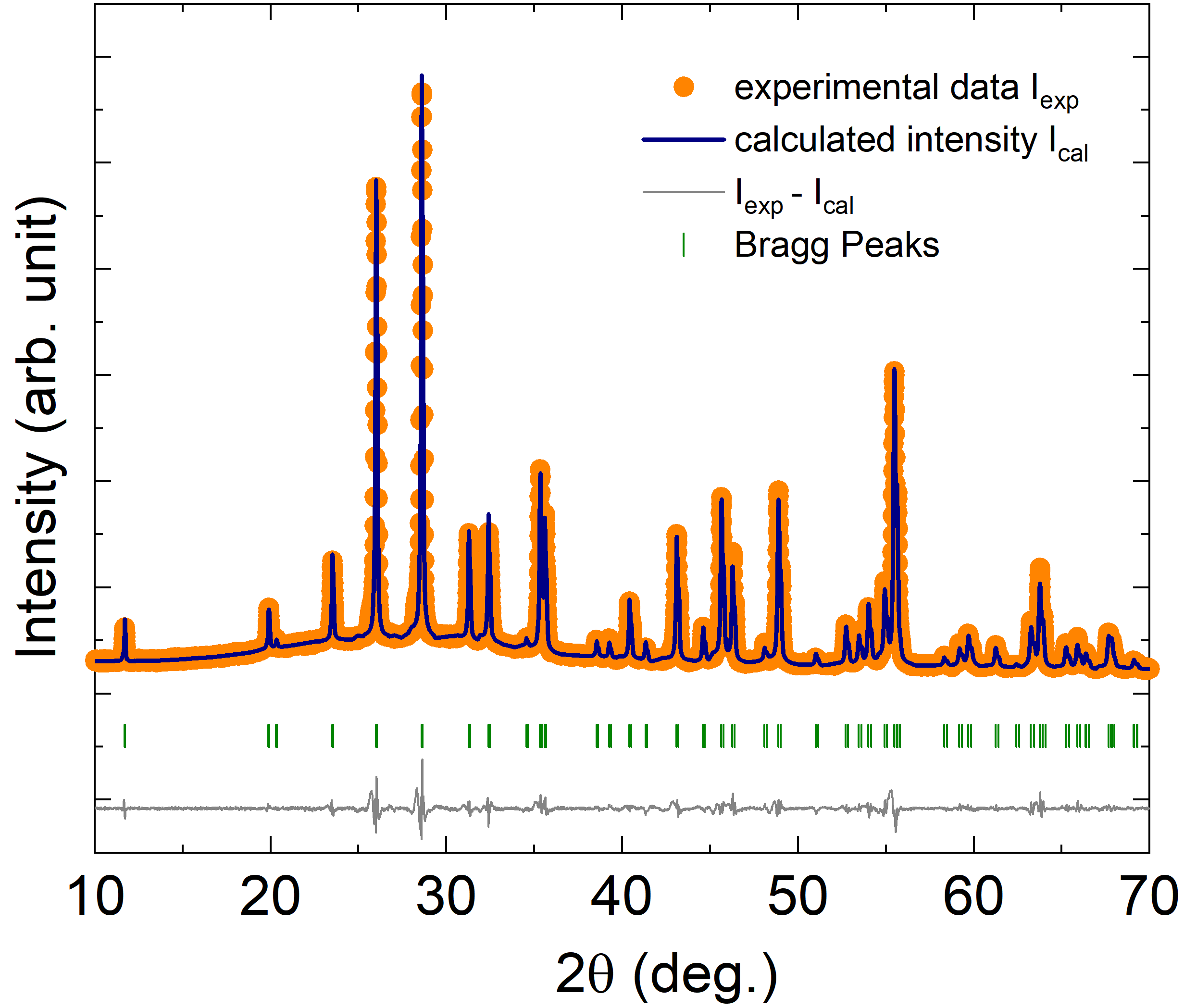}
    \caption{The Rietveld refinement of XRD data of \pbwo\ taken at room temperature. The orange circles, blue solid lines, olive vertical bars, and gray solid lines, are the observed data, calculated data, Bragg peaks, and difference of observed and calculated data of \pbwo, respectively.}
    \label{fig:xrd}
\end{figure}

A polycrystalline sample of \pbwo\ was prepared using a solid-state reaction method. The starting materials Pr$_6$O$_{11}$ (Alfa Aesar, 99.999~\%), H$_3$BO$_3$ (Alfa Aesar, 98~\%), and WO$_3$ (Alfa Aesar, 99.998~\%) were taken in a stoichiometric ratio~\cite{Nagl2024,Krutko2006}. The reagent H$_3$BO$_3$ is highly volatile; therefore, we added 5~\% excess to maintain proper stoichiometry in the sample. Pr$_6$O$_{11}$ was preheated at 900~°C for 12 hours to remove moisture and carbonates prior to use. All the reagents were thoroughly mixed to ensure better homogeneity. An alumina crucible was used to anneal the pellets in the temperature range of 700-900~°C, with several intermediate grindings. X-ray diffraction (XRD) was recorded on the polycrystalline sample of \pbwo\ at room temperature using a benchtop PANalytical diffractometer with Cu K$\alpha$ radiation ($\lambda= 1.541$~\AA). The Rietveld refinement of the XRD pattern at room temperature aims to confirm the phase purity and crystallographic parameters employing FullProf software~\cite{RODRIGUEZCARVAJAL199355}. The Rietveld refinement reveals that the samples used in this study are of high quality, and the refinement parameters are consistent with those reported earlier. Figure~\ref{fig:xrd} depicts the corresponding refinement pattern at room temperature, and the associated refinement parameters are documented in Table~\ref{tab:xrd}.

\begin{table}[htb]
    \caption{The crystallographic parameters determined from the Rietveld refinement of XRD data of \pbwo\ recorded at room temperature (see Fig.~\ref{fig:xrd}). (Space group: $P6_3$
Lattice parameters : $a = b = 8.7315(1)$~\AA, $c = 5.5236(1)$~\AA , $\chi^2=5.2$, $R_{\rm wp} = 4.8$~\%, $R_{\rm p} = 3.31$~\%, and $R_{\rm exp} = 2.09$~\%.)
}
    \centering
    \begin{tabular}{c c c c c c }
\hline\hline
Atom &	Wyckoff &	$x$	&$y$&$	z$&	Occupancy\\
 &	 Pos.&		&&&	\\
\hline
Pr&	6c	&0.0828(1)	&0.7264(2)	&0.365(2)&	1\\
B	&2a	&0.0000&	0.0000&	0.0000	&1\\
W	&2b	&0.3333	&0.6666	&0.893(2)&	1\\
O1	&6c	&0.044(1)&	0.891(1)&	1.012(3)	&1\\
O2	&6c	&0.094(2)&	0.482(1)&	0.165(2)	&1\\
O3	&6c&	0.196(1)	&0.439(1)	&0.642(2)&	1\\
 \hline\hline
\end{tabular}
\label{tab:xrd}
\end{table}

%\pbwo\ crystallizes in the hexagonal space group P63. Pr3BWO9 constitutes a distorted kagomé lattice in the ab-plane, stacked along the c-axis. The distorted kagomé lattice features triangles of two different Pr-Pr bond lengths  one with 4.312 Å and other with 4.886 Å, arranged in an alternating pattern. This suggests that Pr3+ forms a complex exchange network in this rare-earth kagomé magnet as shown in  the inset of Fig.1 [1-3]. [3] is not on \pbwo. Intro!

Magnetic measurements were performed on pressed pellets of $m=6.6(2)$~mg in a MPMS3 magnetometer (Quantum Design) in the temperature regime of 1.8 to 350~K. For measurements down to 400~mK, the MPMS3 was equipped with the iQuantum ${}^3$He setup. The static magnetic susceptibility $\chi =M/B$ has been obtained by varying the temperature using field cooled (FC) and zero field cooled (ZFC) protocols where the sample was cooled in either the external measurement field or the field was applied after cooling to the lowest temperature, respectively. Isothermal magnetization $M(B)$ has been measured in fields up to $B=7$~T. AC magnetization measurements were conducted in the temperature range from 1.8 to 60~K, with 5 - 7~Oe ac excitation fields, up to 5~T dc magnetic fields and at frequencies ranging from 10 to 800~Hz. These measurements were conducted using the ac option of the MPMS3 on a sample with a mass of $m=2.3(1)$~mg. 

Pulsed-field magnetization was measured up to 60~T at Helmholtz Zentrum Dresden-Rossendorf (HLD) by an induction method using a coaxial compensated pick-up coil system~\cite{skourski2011high}. The pulse raising time was 7~ms. Each measurement of the sample was followed by recording the background without a sample under identical conditions, to compensate for background contributions. The pulsed-field magnetization data were calibrated using static magnetic field magnetization data obtained by means of the MPMS3. The field dependence of magnetization obtained by the two methods is in good agreement which confirms the accuracy of our experiments.

To measure the specific heat capacity of the samples, a relaxation method was employed using the heat capacity option of the Physical Property Measurement System (PPMS-14) by Quantum Design. The measurements were carried out in the temperature regime from 1.8 to 200~K on a sample with a mass of 6.6(2)~mg under applied magnetic fields up to 14~T. 

%%%%%%%%%%%%%%%%%%%%%%%%%%%%%%%%%%
\section{Results and Discussion}
%%%%%%%%%%%%%%%%%%%%%%%%%%%%%%%%%%

\subsection{Dc, ac and pulsed field magnetization}

\begin{figure}[b]
    \centering
    \includegraphics[width=\columnwidth]{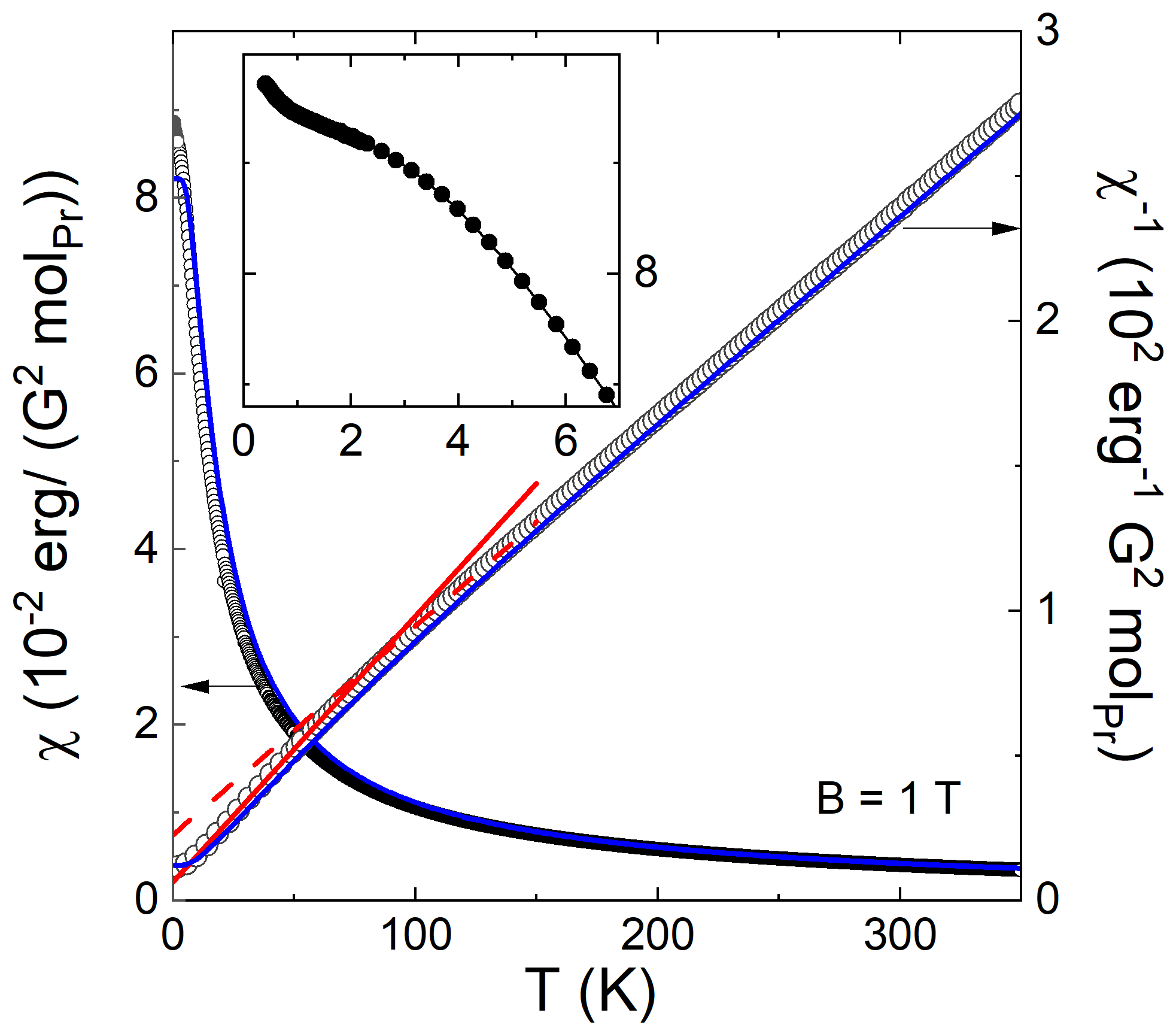}
    \caption{Temperature dependence of the static magnetic susceptibility $\chi = M/B$ and its inverse $\chi^{-1}$ (with reduced number of data points for visibility) measured at $B=1$~T. The dashed (solid) red line represents high (low) temperature Curie-Weiss fits to the data (see the text). The blue line depicts the simulation of a non-interacting single ion model for Pr$^{3+}$ using the crystal field parameters obtained in Ref.~\cite{Nagl2024}. The inset shows the enlargement of the low-temperature regime.}
    \label{fig:Pr_Xdc_CW}
\end{figure}

%\begin{figure}[b]
%    \centering
%    \includegraphics[width=\columnwidth]{CW5.png}
%    \caption{Temperature dependence of the static magnetic susceptibility $\chi = M/B$ and its inverse $\chi^{-1}$ measured at $B=1$~T. The dashed (solid) red line represents high (low) temperature Curie-Weiss fits to the data (see the text). The yellow line is a fit to $\chi^{-1}$ according to Eq.~\ref{CWthree} (see §~\ref{sec:HC}). The inset shows the enlargement of the low-temperature regime.}
%    \label{fig:Pr_Xdc_CW}
%\end{figure}

\begin{figure}[tb]
\centering
\includegraphics[width=\columnwidth]{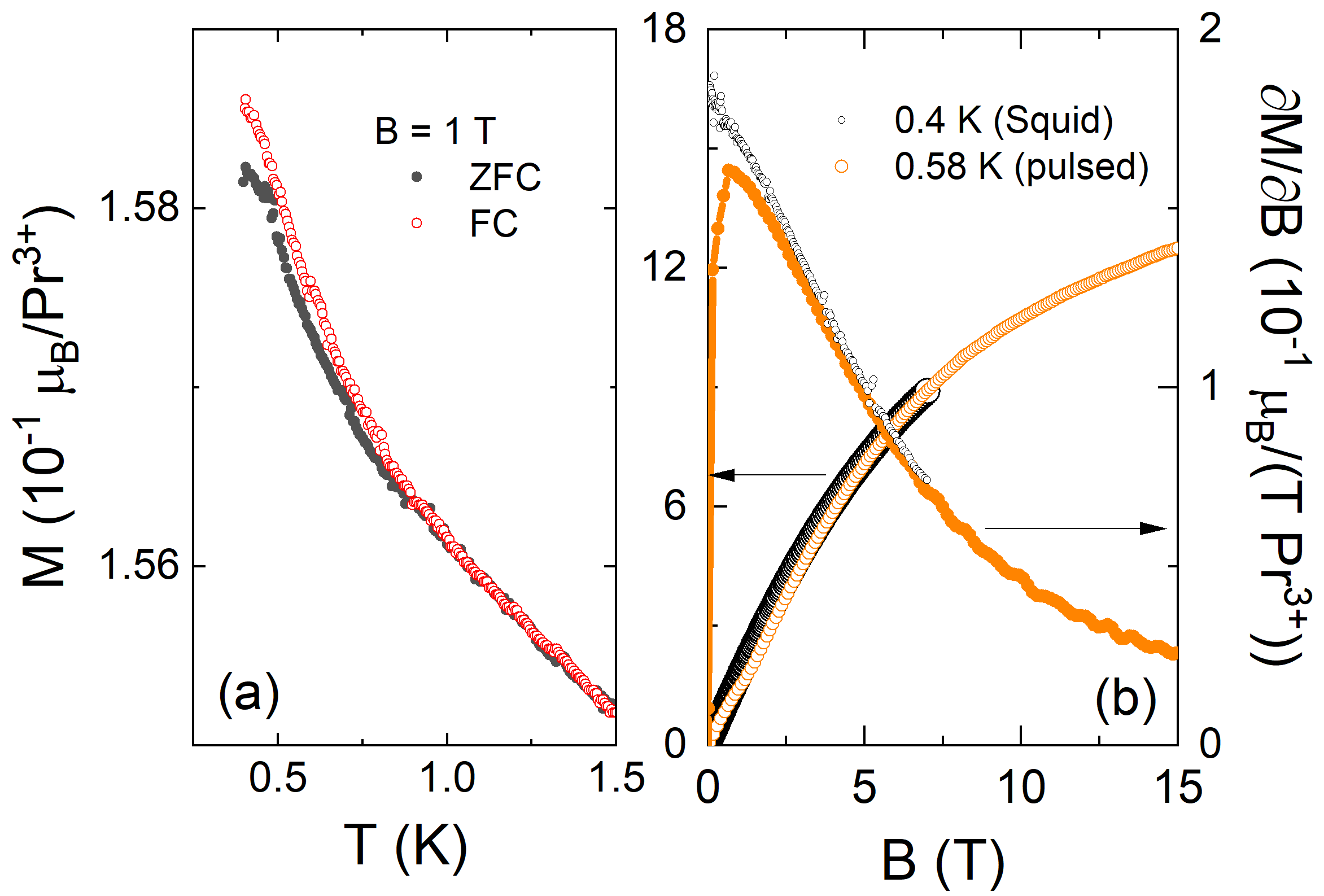} %scale=0.5
    \caption{(a) Temperature dependence of the static susceptibility, at $T<1.5$~K, and (b) field dependence of the magnetization and of the differential magnetic susceptibility $\partial M/\partial B$, at low temperature. Black open (orange filled) data points refer to static (pulsed) field measurements. }
    \label{fig:MBT}
\end{figure}

The static magnetic susceptibility of \pbwo\ at $B=1$~T, shown in Fig.~\ref{fig:Pr_Xdc_CW}, decreases upon heating and does not exhibit any pronounced anomalies in the whole temperature regime 0.4~K~$\leq T\leq$~350~K under study. In particular, the data do not show any signature of long-range magnetic order down to 400~mK which is in agreement with recent specific heat, inelastic neutron and $\mu$SR data~\cite{zeng2021local,Nagl2024}. In the literature, absence of long-range magnetic order down to at least 90~mK~\cite{zeng2021local} is reported.
%Furthermore, no bifurcation between ZFC and FC measurements is found down to at least 470~mK which excludes the presence of a spin-glass state in this regime (see Fig.~\ref{fig:MBT}a).
The low-temperature $\chi(T)$ indicates saturation below $\simeq 5$~K superimposed by a small upturn starting from $\sim 0.7$~K. Based on the observation of a weak FC/ZFC bifurcation in the same temperature regime (Fig.~\ref{fig:MBT}a), the latter is likely not associated with quasi-free moments from putative defects but might rather be related to a spin freezing phenomenon typically observed in spin glass materials~\cite{nagata1979low,petrenko2011titanium,krey2012first,Elghandour2024}. This scenario is in qualitative agreement with the observation of a peak at $B = 0.7$~T and $M/M_{\rm sat} \sim 1/5$ in the pulsed field $\partial M/\partial B$ data at $T\simeq 0.5$~K which is absent in the static field measurements (Fig.~\ref{fig:MBT}b). Our results, thus, indicate that a tiny fraction of \prion\ moments start to freeze below $\sim 0.7$~K while the bulk of the spins maintains a dynamic state.

\begin{figure}[t]
\label{fig:Pr_Xac}
        \includegraphics[width=\columnwidth]{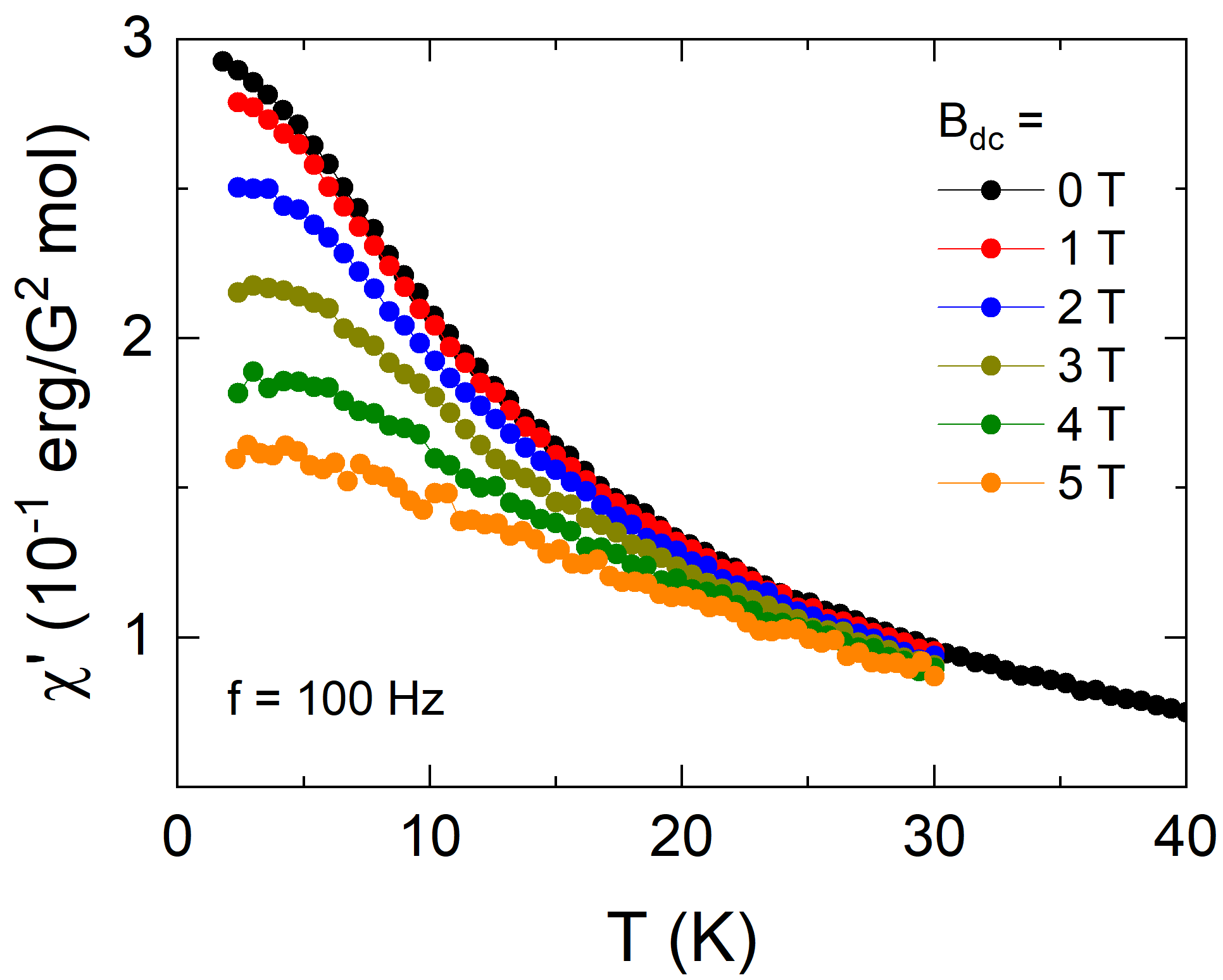}
        \caption{Temperature dependence of the real part of the ac magnetic susceptibility $\chi'$ measured at $f=100$~Hz and $B_{\rm ac} = 0.5$~mT, at different static magnetic fields. Changing the frequencies in the range of 10~Hz~$\leq f \leq$~800~Hz yields no changes of the signal (data not shown).}
\end{figure}

At high temperatures, $\chi(T)$ exhibits a Curie-Weiss-like behavior. Fitting the data between 150~K~$\leq T\leq$~350~K by means of the Curie-Weiss model yields an effective magnetic moment of $p_{\rm eff}=3.33(3)$~\mb\ and the Weiss temperature $\Theta_{\rm high} = -31(2)$~K in agreement with previous reports~\cite{ashtar2020new}. While the negative sign of $\Theta_{\rm high}$ indicates dominant antiferromagnetic interactions, $p_{\rm eff}$ is smaller than the theoretical free ion value of 3.58~\mb\ for Pr$^{3+}$. As will be shown when discussing specific heat data, we attribute this deviation from the free ion value to the presence of crystal field levels of \prion\ which are not (fully) thermally populated at the highest measurement temperature of 350~K. We in particular note that employing the non-interacting single-ion crystal-field model from Ref.~\cite{Nagl2024} yields a good description of our magnetic susceptibility data (cf.~Fig.~\ref{fig:Pr_Xdc_CW}) which implies very weak or nearly vanishing magnetic interaction. Note, that an estimate of bare magnetic dipolar interaction energy yields $E_{\rm dip}< 0.2$~K. We conclude that the data only seemingly resemble a Curie-Weiss-like behavior at intermediate temperatures but actually are dominated by the thermal population of excited crystal field levels. Upon cooling, a second quasi-linear regime is observed in $\chi^{-1}$ below $\sim 40$~K (see Fig.~\ref{fig:Pr_Xdc_CW}). Fitting the data in the temperature regime 10~K~$\leq T\leq$~25~K results in $p_{\rm eff}=2.95(1)$~\mb\ and $\Theta_{\rm low} = -6.8(2)$~K which are consistent with those reported in Refs.~\cite{ashtar2020new,zeng2021local,Nagl2024}. This behavior reflects the subsequent depopulation of \prion\ crystal field levels so that neither $\Theta_{\rm high}$ nor $\Theta_{\rm low}$ can be taken as a reliable measure of the strength of magnetic interaction and, in contrast to previous reports, we suggest that it should not be used to estimate the frustration parameter. 

The ac magnetic susceptibility of \pbwo\ shows neither a dissipative signal in the imaginary part $\chi''$ (data not shown) nor any frequency dependence in the real part within the studied temperature and frequency range, i.e., at $T\geq 1.8$~K and 10~Hz~$\leq f \leq$~600~Hz. Consistent with the static susceptibility shown in Fig.~\ref{fig:Pr_Xdc_CW}, the real part of ac susceptibility $\chi '$ increases monotonically upon cooling. The frequency independence implies that at $T\geq 1.8$~K, which is well above the temperature regime where the tiny bifurcation is observed in $\chi_{\rm dc}(T)$ (cf. Fig.~\ref{fig:MBT}a), \pbwo\ either displays no glassy behavior or the characteristic spin-fluctuation rate in \pbwo\ is beyond the kilohertz frequency regime. The latter scenario is suggested by NMR studies (MHz frequencies) revealing the existence of short-ranged collective spin excitations~\cite{zeng2021local}. The application of external dc magnetic fields barely affects the behavior of $\chi'(T)$ as seen in Fig.~\ref{fig:Pr_Xac} and, in particular, does not induce clear anomalies. Note that this is in stark contrast to the findings in \nbwo~\cite{Yadav2025} thereby demonstrating strong differences in the spin relaxation rates of the two systems. One may speculate that spin relaxation rates in \pbwo\ are much higher than those in \nbwo\ where already moderate magnetic fields of a few Tesla induce slow spin relaxation at tens of Kelvin while spin freezing appears at 0.6~K.

\begin{figure}[tb]
\centering
\includegraphics[width=\columnwidth]{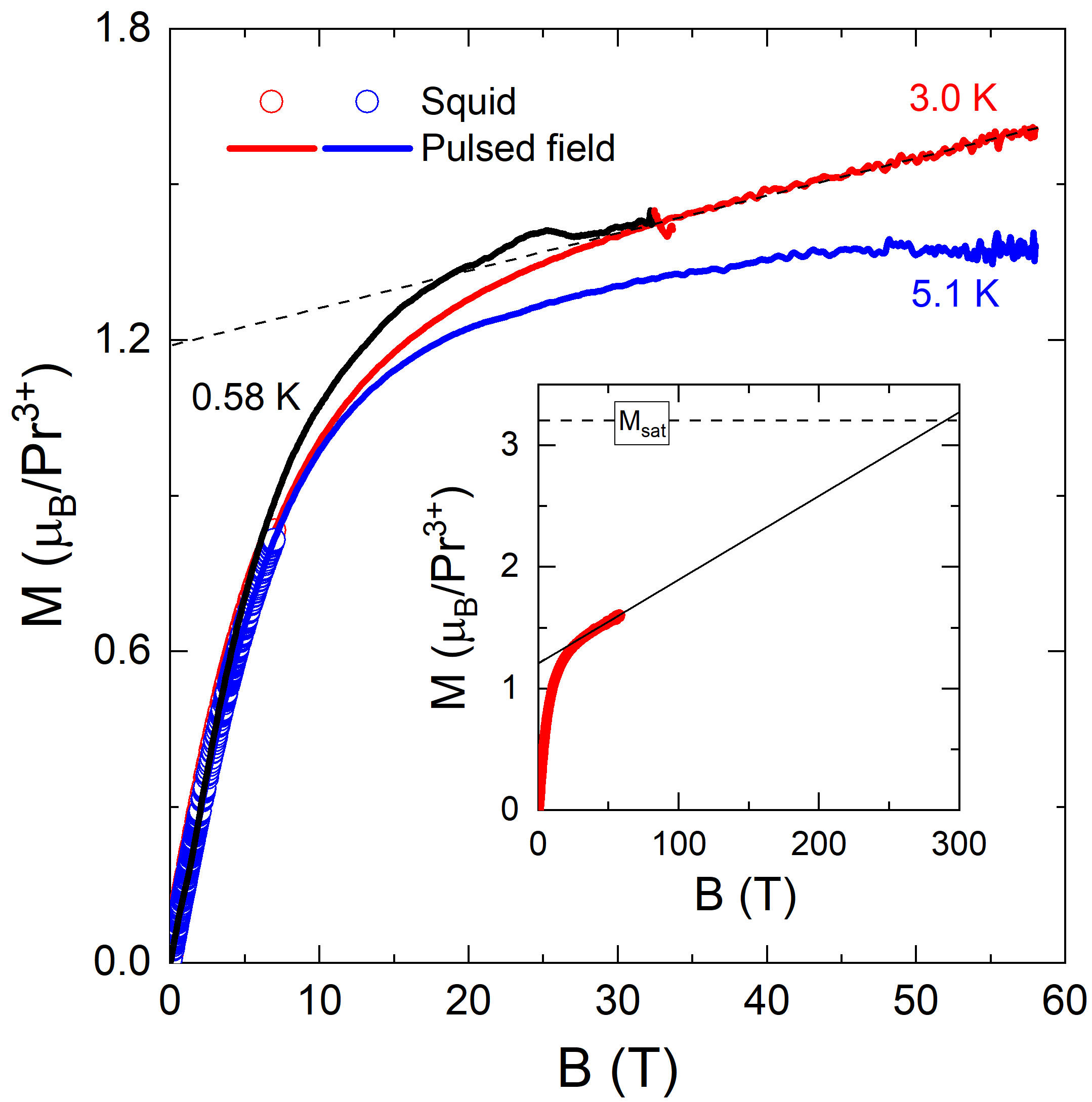} %scale=0.5
    \caption{Magnetic field dependence of the static and the pulsed field magnetization of \pbwo\ measured at different temperatures in up-sweeps of magnetic fields up to 7~T and 58~T, respectively. Inset: $M(T=3~{\rm K},B)$ extrapolated up to the saturation value $M_{\rm sat}=3.2$~\mbpr .}
    \label{fig:Pr_DC_pulsed}
\end{figure}

The magnetization at low temperature increases quasi-linearly up to $\simeq 2$~T and shows right-bending at higher magnetic field without any clear additional features (Figs.~\ref{fig:MBT} and \ref{fig:Pr_DC_pulsed}). At the maximum applied magnetic field of 58~T and $T=3$~K, $M$ amounts to 1.60(2)~\mb/Pr$^{3+}$. Extrapolating the linear-in-field regime which is observed at $B>30$~T to zero magnetic field yields 1.2~\mbpr . In a pure paramagnet, this value reflects the saturation magnetization of the powder averaged ground state singlet. Considering an almost uniaxial crystal field and $g_J=4/5$ of the free ion, and powder averaging, this suggests an Ising ground state with $|m_J|=4$. Our data also allow us to roughly estimate the anisotropy field by linearly extrapolating the pulsed field magnetization up to the full saturation magnetization of 3.2~\mbpr\ associated with the complete alignment of \prion\ moments of the powder sample along the magnetic field direction: following the dashed line in Fig.~\ref{fig:Pr_DC_pulsed} suggests a saturation field of $B_{\rm sat}\simeq 285$~T (inset of Fig.~\ref{fig:Pr_DC_pulsed}). We hence conclude the anisotropy energy $E_a/k_{\rm B}=g_J\mu_{\rm B}m_JB\simeq 950$~K ($E_a \simeq 82$~meV). In summary, our pulsed field magnetization data suggest a strong uniaxial crystal field and an Ising ground state in \pbwo\ characterized by a (mainly) $m_J=-4$ ground state singlet, in agreement to Ref.~\cite{Nagl2024}. This finding implies the dominance of anisotropy energy over magnetic exchange interactions which is consistent with the observed absence of a 1/3 magnetization plateau (in the pulsed field data) or a 1/9 plateau (which would also show up in the MPMS3-static field data) which are typically observed in Heisenberg-like kagome antiferromagnets (see, e.g.~Refs.~\cite{Zhitomirsky2002,Nishimoto2013,Schnack2018,Okuma2019,Jeon2024}). 

%Although these values are far from the theoretical saturation magnetization $M_{sat} = 3.2$~\mb/\prion\ expected for free \prion\ ions, they agree well with the values reported for distorted Kagom\'e system Pr$_3$Ga$_5$Si$_3$O$_{14}$~\cite{hoch2016pulsed} from pulsed-field magnetization data, and with other \prion-based frustrated magnets Pr$_3$M$_2$Sb$_3$O$_{14}$ (M = Mg, Zn)~\cite{sanders2016re3sb3zn2o14,sanders2016synthesis}. Note, the quasi-static magnetization of \PBWO\ polycrystal sample reported in~\cite{ashtar2020new} gives $M_{sat} = 1.2 $~\mb/\prion\ at the highest field of 14~T and measurement at 2~K.

\subsection{Heat Capacity}\label{sec:HC}

\begin{figure}[bt]
    \centering
    \includegraphics[width=\columnwidth]{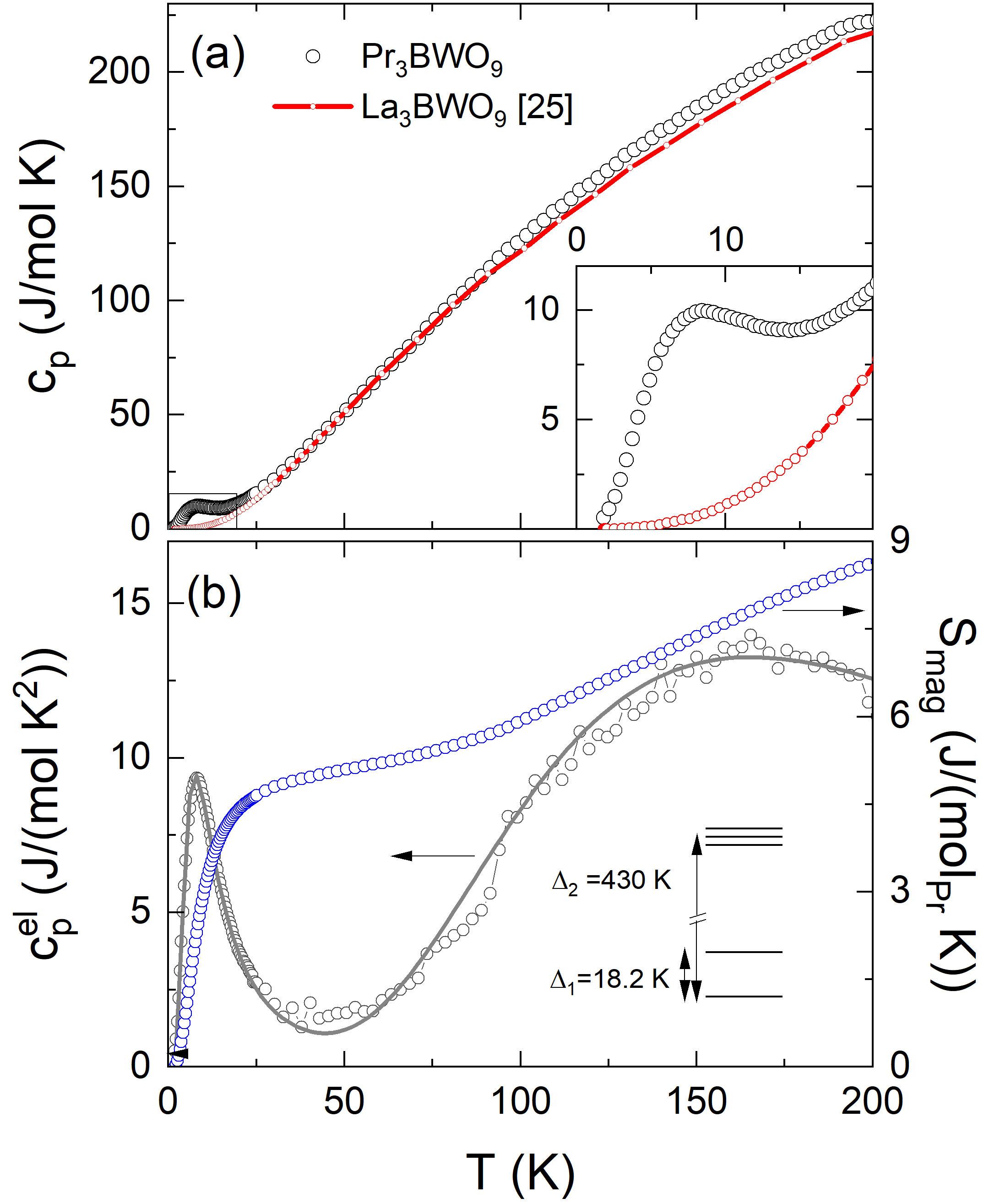} %scale=0.5
    \caption{Temperature dependence of (a) the specific heat of \pbwo\ and \lbwo\  measured at $B=0$~T; the latter data are taken from Ref.~\cite{Flavi}. Inset: Enlargement of the low-temperature regime. (b) Non-phononic specific heat \cpel~ of \pbwo\ as  obtained by the subtraction of \cpp\ \lbwo\ from the one of \pbwo, and calculated magnetic entropy $S_{\rm mag} =\int (c_{\rm p}^{\rm el}(T)/T) dT$. The line is a fit according to an extended three-level Schottky model (see Eq.~3) with $\Delta_1=18.2$~K, $\Delta_2=430$~K, and level degeneracies $g_0=1$, $g_1=1$, and $g_2=3$ (see the text).}
    \label{fig:Pr_cp_smag}
\end{figure}

\begin{figure}[htb]
    \centering
    \includegraphics[width=\columnwidth]{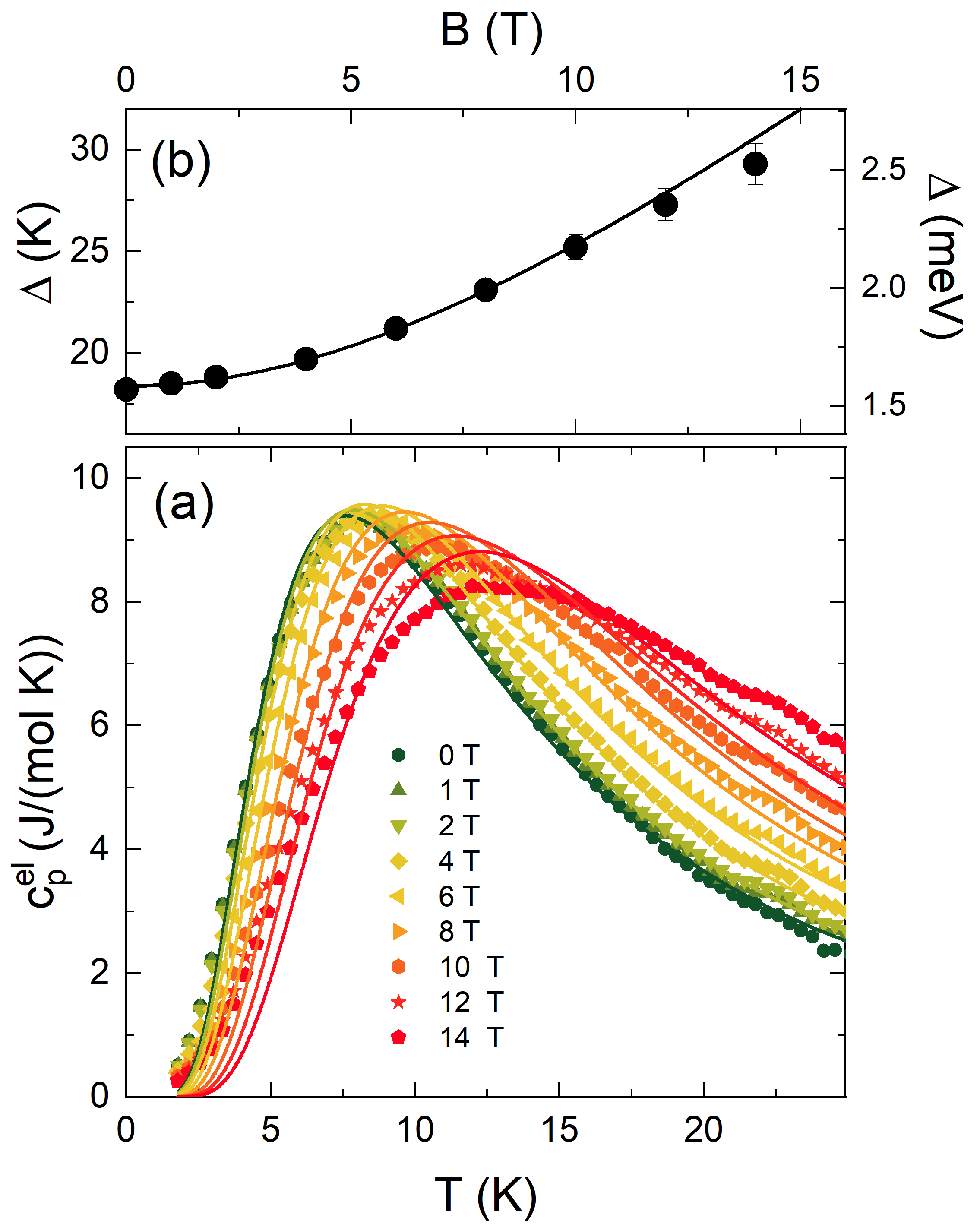} %scale=0.5
    \caption{(a) Temperature dependence of the electronic specific heat \cpel\     
    of \pbwo\ for different applied magnetic fields. The solid lines represent the fits to the two-level Schottky function according to Eq.~\ref{eq_Schottky}. (b) Magnetic field dependence of the energy gap $\Delta_1$ obtained from the fitting \cpp\ with the Schottky model. The line illustrates the cubic field dependence of $\Delta$. }
    \label{fig:Pr_Sch_Fits}
\end{figure}

The electronic contribution to the specific heat capacity of Pr$_3$BWO$_9$ is estimated by subtracting the phononic contribution of the isostructural non-magnetic La$_3$BWO$_9$ as reported for $T<200$~K in Ref.~\cite{Flavi}. In order to account for the different molar masses, the scaling factor of 0.97 (see Ref.~\cite{tari2003specific}) has been applied. The resulting \cpel\ is presented in Fig.~\ref{fig:Pr_cp_smag}b and features a peak centered at around 5.7~K which is well separated from an additional, much broader hump at higher temperatures. The electronic specific heat is attributed to electronic or, specifically, magnetic degrees of freedom. The entropy changes in the temperature regime 1.8 -- 40~K, i.e., $\Delta S_{\rm mag} =\int_{1.8~{\rm K}}^{40~{\rm K}} (c_{\rm p}^{\rm el}(T)/T) dT$ associated with the low-temperature peak amount to 4.9~\jmk . Adding the entropy changes of 0.11~\jmk\ from the heat capacity data at 0.09~K~$\leq T\leq$~1.8~K reported in Ref.~\cite{zeng2021local}, $S_{\rm mag}$ amounts to only 30\% of the full expected magnetic entropy of $R\ln(8) = 18.3$~\jmk~for $J = 4$, but it represents 86\% of the expected value ($R\ln{2} = 5.76$~\jmk) of a two-level system with $J_{\mathrm {eff}} = 1/2$, where $R$ is the universal gas constant. We note that in the isomorphic \nbwo\ the magnetic entropy changes are, too, well described by a two-level system~\cite{Flavi,Yadav2025} but in the Nd-based sister compound the magnetic entropy is released below $T = 3$~K, i.e., at one tenth of the saturation temperature of the lowest energy gap found in \pbwo .

The broad anomaly in the electronic specific heat centered around 8~K shown in Fig.~\ref{fig:Pr_cp_smag} resembles a Schottky-like behavior and can be attributed to electronic excitations within the split $J = 4$ ground state manifold of the Pr$^{3+}$ ions~\cite{zeng2021local,Nagl2024}. The Schottky nature is clearly corroborated by the observed magnetic field dependence which shows a gradual shift of the peak position to higher temperatures, peak broadening and peak height suppression upon application of external magnetic fields (cf.~Fig.~\ref{fig:Pr_Sch_Fits}a). The low-temperature peak is associated with the energy gap between the lowest crystal field-split energy levels and, hence, can be used to investigate, e.g., the gap size and magnetic field dependence, and the degeneracies of the associated energy levels. For the non-Kramer's ion Pr$^{3+}$ with $J = 4$, the crystal field effects in \pbwo\ result in nine singlet states, some of which may have accidental or near-degeneracies~\cite{scheie2018crystal,zeng2021local,Nagl2024}. To quantify the energy gap separation between the ground state and the first excited level, the Schottky model for the specific heat capacity of a gapped two-level system is fitted to the data~\cite{tari2003specific}:

\begin{equation}
c_{\text{p}}^{\Delta_1} = n \cdot R \left(\frac{\Delta_1}{T}\right)^2 \frac{g_{0}}{g_{1}} \frac{e^{\frac{\Delta_1}{T}}}{\left[1 + \left(\frac{g_{0}}{g_{1}}\right) e^{\frac{\Delta_1}{T}}\right]^2}.
\label{eq_Schottky}
\end{equation}

Here, $n$ is the concentration of magnetic sites, $R$ is the universal gas constant, $\Delta_1$ is the associated energy gap, and ($g_{\mathrm {0}}/g_{\mathrm {1}}$) is the degeneracy ratio of the two energy levels. Using $n$ and $\Delta_1$ as free fit parameters while fixing $g_{\mathrm {0}}/g_{\mathrm {1}}$ = 1 yields a good description of the experimental data \cpel\ as represented by the solid lines in Fig.~\ref{fig:Pr_Sch_Fits}a. The resulting fit parameters are listed in Table~\ref{Sch_fits}. We note that the best fits are achieved with $g_{\mathrm {0}}/g_{\mathrm {1}}$ kept constant which implies that the magnetic field effect on the levels' degeneracies is negligible. Further, the fits yields essentially field independent $n\simeq 2.6$  which indicates that only 85\% of the Pr$^{3+}$ ions contribute to the Schottky specific heat. This agrees well with the observed entropy changes of $\sim0.85\times R\ln{2}$. %In the same temperature region, the low-temperature Curie-Weiss fit yields $p=2.95$~\mbpr\ as described above which agrees to the same fraction of the theoretical value. SEE THE NEW MODELLING. 
%{\rk The fits become slightly worse at higher fields, which possibly arises from higher energy levels (most likely $m_J = -3$) starting to become relevant. XX Not sure that this is correct for only 14 T fields.} {\jan I mean, the fits become gradually worse. This would be my first guess.}

\begin{table}[tb]
%\begin{minipage}{\columnwidth}
    
    \caption{Parameters describing the electronic specific heat \cpel\ obtained by fitting the data to a two-level Schottky (Eq.~\ref{eq_Schottky}) model at different applied magnetic fields (see Fig.~\ref{fig:Pr_Sch_Fits}). $\Delta_1$ is the splitting gap and $n$ is the concentration of Schottky centers.  }
    \centering
    \begin{tabular}{c c c  }
\hline\hline
$B$ (T) & $n$ & $\Delta_1$ (K)  \\
 \hline
 0 & 2.57(1) & 18.2(1)  \\
 1 & 2.58(1) & 18.5(1)  \\
 2 & 2.57(1) & 18.9(1)  \\
 4 & 2.62(2) & 19.8(2)  \\
 6 & 2.61(2) & 21.2(2)  \\
 8 & 2.58(1) & 23.1(2)  \\
 10 & 2.54(2)& 25.2(3)  \\
 12 & 2.48(2)& 27.3(4)  \\
 14 & 2.41(1)& 29.5(6)  \\
 \hline\hline
\end{tabular}
\label{Sch_fits}
\end{table}

The magnetic field dependence of the obtained $\Delta$ is plotted in Fig.~\ref{fig:Pr_Sch_Fits}b. It is well described by the expected quadratic behavior

\begin{equation}
    \Delta_1(B)=\sqrt{(\Delta_{1}(0))^2+(g\mu_{\rm B}B/k_{\rm B})^2}
\end{equation}

with $\Delta_{1}(0)=\Delta_1(B=0~{\rm T})$ and 
the effective $g$-factor, $g$. The best fit yields $\Delta_1(0)=18.4$~K (i.e., $1.6$~meV). This value agrees very well to the excitation gap of 1.62(1)~meV found in inelastic neutron scattering~\cite{Nagl2024} but is considerably larger than the excitation gap of 8.0~K (14.1~K) found in NMR for $B||c$ ($B\perp c$) axis~\cite{zeng2021local}. The field dependence is best described by $g\simeq2.6$ which is in very good agreement with the powder average of the Ising-like low-energy $m_J$ states as described by the crystal field parameters in Ref.~\cite{Nagl2024}. The observed quadratic dependence $\Delta_1(B)$ qualitatively resembles the findings of Ref.~\cite{Nagl2024} while Ref.~\cite{zeng2021local} reports a linear field dependence of $\Delta$. Note, that the pulsed field magnetization data in Fig.~\ref{fig:Pr_DC_pulsed} do not indicate field-induced level crossing, e.g., closing of the gap $\Delta_1$. Such a closing of the gap would be observed in the high-field measurements if, e.g., the first excited singlet was characterized by larger (i.e., more negative) effective $m_J$ as compared to the ground state. This observation is in  agreement to the Ising-like scenario.

The second hump in \cpel\ appearing around 160~K indicates the thermal population of one or more energetically higher-lying crystal field levels of \prion . We hence extend the Schottky model Eq.~\ref{eq_Schottky} which is used to describe the entropy changes associated with the thermal population of the crystal field levels to a three-level model~\cite{Souza2016}):

\begin{widetext}
\begin{equation}
c_{\rm p}^{\Delta_{1}/\Delta_{2}}=nR\times\frac{g_1g_0\Delta_1^2g_1e^{-\Delta_1/T}+g_0g_2\Delta_2^2g_1e^{-\Delta_2/T}+g_1g_2e^{-(\Delta_1+\Delta_2)/T}[\Delta_1(\Delta_1-\Delta_2)+\Delta_2(\Delta_2-\Delta_1)]}{T^2[g_0+g_1e^{-\Delta_1/T}+g_2e^{-\Delta_2/T}]^2}.\label{eq:two}
\end{equation}
\end{widetext}

Here, the energies of the three lowest states are $0$, $\Delta_1$ and $\Delta_2$ while $g_i$ ($i=0,1,2$) is the degeneracy of the respective level. Fitting the experimental data by means of Eq.~\ref{eq:two} yields an excellent description of the data. Using $n=2.6$ obtained from the low-temperature Schottky peak, \cpel\ is well described by $\Delta_1=18.2$~K and $\Delta_2=430$~K (see the grey line and the sketch in Fig.~\ref{fig:Pr_cp_smag}b. While the degeneracies of the ground state and the first excited state are $g_0=1$ and $g_1=1$, the second excited state exhibits $g_2=3$. These results are in reasonable agreement with crystal field splittings suggested by a point-charge model in Ref.~\cite{Nagl2024} which predicts a ground state quasi-doublet while the next excited states are well isolated singlets at energies of $\Delta_2=429$~K (37.0~meV), $\Delta_3=494$~K (42.6~meV), and $\Delta_4=673$~K (58.0~meV). However, as clearly seen in the experimental data, the predicted gap values neither are consistent with our specific heat nor do we find signatures for well separated singlets at the predicted energies $\Delta_2, \Delta_3$ and $\Delta_4$. In contrast, the experimental data imply three quasi-degenerated levels. Quantitatively, our data support a slightly smaller value of $\Delta_2$ than predicted by the point-charge model and clearly disagree with the prediction for $\Delta_4$.

\subsection{Summary}

In summary, we report detailed dc and ac magnetization studies of the frustrated kagome system Pr$_3$WBO$_9$ in a wide magnetic field and temperature regime, i.e., up to 60~T in pulsed magnetic fields and up to 7~T in static fields and down to 0.4~K. This is supported by specific heat studies up to 14~T. Our data confirm a highly frustrated Ising-like system the low-temperature properties of which are governed by two moderately split singlets ($\Delta_1\simeq 18$~K) forming a quasi-doublet well separated from all other crystal field levels. The next excited states are identified at $\Delta_2\simeq 430$~K. The low-energy gap $\Delta_1$ increases quadratically in external magnetic fields and the field dependence is governed by the effective $g$ factor 2.6. No distinct glassy behavior is observed in our dc and ac magnetization studies down to 400~mK. The magnetic ground state is further investigated by magnetization studies up to 60~T confirming the Ising-like paramagnetic nature of the magnetic ground state which is mainly characterized by $m_J= -4$ with anisotropy energy $E_a/k_{\rm B}\simeq 950$~K. The magnetic susceptibility is well described by an isolated single-ion model. Our results hence add \pbwo\ to the growing family of Ising-like yet disordered complex lanthanide-based magnets. Additional studies of magnetic excitations and magnetic exchange interactions on high-quality single crystals will be needed to further elucidate the quantum Ising ground state and associated excitations in this class of frustrated quantum magnets.

%extended Curie-Weiss model. A multi-level Curie-Weiss-like model using the energy gaps obtained by the specific heat analysis yields the Weiss temperature $\Theta \simeq -18$~K which confirms predominant antiferromagnetic interaction of the \prion\ moments and suggest a high frustration ratio $f = |\Theta|/T_{\mathrm{N}} > 200$ and adds \pbwo\ to the growing family of Ising-like yet disordered complex lanthanide-based magnets.
%{\rk Further studies on the high-quality single crystals may provide interesting insights into the quantum Ising state state and associated magnetic excitations in this class of frustrated quantum magnets.
%XX (1) What is particularly 'quantum' in the CF split ground state? In this respect I also do not understand Zheludev. (2) I still think it's not a helpful outlook since this work has to a very large extend already been done by Zheludev and hence weakens our work. }\\

\begin{acknowledgements}
Support by Deutsche  Forschungsgemeinschaft (DFG) under Germany’s Excellence Strategy EXC2181/1-390900948 (The Heidelberg STRUCTURES Excellence Cluster) is gratefully acknowledged. A.E. acknowledges funding through the DAAD GSSP program. We acknowledge the support of the HLD at HZDR, a member of the European Magnetic Field Laboratory. J.A. acknowledges support by the IMPRS-QD Heidelberg. P.K. acknowledges the funding by the Science and Engineering Research Board, and Department of Science and Technology, India through Research Grants.
\end{acknowledgements}

\bibliography{ref}

%\section{Appendix}
%\label{sec:appendix}

%\renewcommand{\thefigure}{S\arabic{figure}}
%\setcounter{figure}{0}  

%For convenience of the reader, we cite the adopted Eq.~6 of Ref.~\cite{Souza2016} which we used to describe the electronic specific heat of the three-level system in Sec.~\ref{sec:hc}:

%\begin{widetext}
%\begin{equation*}
%c=nR\frac{1}{T^2}\frac{g_1g_0\Delta_1^2g_1e^{-\Delta_1/T}+g_0g_2\Delta_2^2g_1e^{-\Delta_2/T}+g_1g_2e^{-(\Delta_1+\Delta_2)/T}[\Delta_1(\Delta_1-\Delta_2)+\Delta_2(\Delta_2-\Delta_1)]}{[g_0+g_1e^{-\Delta_1/T}+g_2e^{-\Delta_2/T}]^2}.
%\end{equation*}
%\end{widetext}

%Here, the energies of the three lowest states are $0$, $\Delta_1$ and $\Delta_2$. $g_i$ ($i=0,1,2$) is the degeneracy of the respective level, $R$ is the universal gas constant and $n$ accounts for the concentration of paramagnetic centers.

\end{document}